# Simulation of Kinematics of Special Theory of Relativity

V. N. Matveev and O. V. Matvejev

Joint-Stock Company "Sinerta"
Savanoriu pr., 159, Vilnius, LT-03150, Lithuania
E-mail: matwad@mail.ru

**Abstract**

The principles of the special theory of relativity are extremely simple. A knowledge of the Pythagorean theorem and an ability to perform the simplest algebraic operations are sufficient to be conversant with the kinematics of the special theory of relativity, as well as the time dilation and contraction of the longitudinal dimensions of moving bodies that are associated with relative motion. However, the simplicity of the fundamentals of the theory of relativity are in surprising contrast with the difficulty of the perception and at times the total nonacceptance of the consequences of the special theory of relativity by skeptics based on ordinary common sense. The authors of certain popular books on the theory of relativity explain the existence of this contrast by way of the fact that the common sense of skeptics cut its teeth on a "stark notion of our everyday life".

The special theory of relativity is simulated in this article based on the simplest examples of the movement of barges, shuttles, and boats in an aquatic environment. It has been simulated without rejecting the customary ordinary common sense. In this article, relativistic time and the relativistic effects of Einstein's special theory of relativity – Lorentz contraction, time dilation, relativistic Doppler effects, the Skobeltsyn-Bell effect, and the relativistic addition of velocities – are simulated using elementary methods of classical physics. They have been simulated without rejecting the customary ordinary common sense that was squashed in the past century by the celebration of "mad ideas". Lorentz transformations are obtained. Means for simulating four-dimensional space-time are shown.

**Introduction**

Many people even today, perhaps "through inertia", either are not aware of the importance of the question of clock synchronization or are very poorly informed of its essence in general.

The problematic nature of synchronizing clocks consists of the use of the condition of the equality of the speed of light in opposite directions for clock synchronization in the STR, while it is fundamentally impossible to experimentally verify this equality. In order to measure the speed of light from point $A$ to point $B$, then from point $B$ to point $A$, and then to compare these speeds, it is necessary to have synchronously running clocks at points $A$ and $B$. However, it is only possible to synchronize the clocks at points $A$ and $B$ using the Einstein method by assuming that these velocities are equal even before they are measured. Naturally, after this assumption is made, they also become equal based on the measurement results.

It is also not possible to measure velocity by synchronizing a pair of clocks at point $A$, then moving one of them to point $B$, since the result of the synchronization and measurement of the speeds of light from point $A$ to point $B$ and back, $v_{AB}$ and $v_{BA}$, respectively, is dependent upon the speed at which the clocks are transported from one point to the other. When synchronizing clocks via the transfer technique, if the clocks being transported are transferred at different speeds in different instances, the $v_{AB}$ and $v_{BA}$ velocity measurement results will then be different in different instances. For example, after a clock is transferred from $A$ to $B$ at a velocity close to the speed of light, the $v_{AB}$ velocity subsequently measured will be arbitrarily great, while the $v_{BA}$ velocity will be arbitrarily close to $c/2$. During such synchronization, the light arrives at point $B$ from point $A$ almost instantaneously, but travels back two times slower than usual. During very slow transfer, the $v_{AB}$ and $v_{BA}$ velocities will be equal to one another.

So what clock transfer speed is "correct"? It is impossible to answer this question.

It should be noted that the problem of the speed of light in one direction is not of topical interest in practice, since the speed of light is actually measured using one solitary clock and a mirror. During this solitary clock method, the time interval between light pulse dispatch to the mirror and the reception of the pulse returned to the initial point after being reflected from the mirror is measured. Velocity is calculated for the doubled distance between the clock and the mirror and the light travel time in the back and forth directions. Strictly speaking, velocity measured in this manner constitutes the average speed in the back and forth directions – this is because the speed there may not equal the speed back. The equality of this average speed of the $c$ constant is an experimental fact.

No clock synchronization problems arise during the measurement of average speed.

It is frequently said that Rømer measured the speed of light in one direction. It may seem strange, but Rømer velocity is also the velocity obtained under the tacit assumption of the equality of the speeds of light in opposite directions. The fact of the matter is that Rømer and Cassini were speculating about the movement of Jupiter's satellites, automatically assuming that the observers' space was isotropic. The Australian physicist Karlov [1] showed that Rømer actually measured the speed of light by implicitly making the assumption of the equality of the speeds of light back and forth.

Poincare examined the proposition of the equality of the speed of light from *A* to *B* and the speed from *B* to *A*, and this proposition in particular became the principal postulate of Einstein's 1905 work [2], although it was not presented in the form of a postulate, but rather in the form of a "definition", which preceded two Einsteinian principals that are often called postulates.

Within the framework of the pre-Einsteinian ethereal world view, it was thought to be natural that in a laboratory moving through the ether at a velocity of *v*, the longitudinal speed of light, $c_{long}$, in the direction of its movement – from point *a* on the "stern" to point *b* on the "bow" – equals *c–v*, while in the direction opposite that of movement (from point *b* to point *a*), *c+v*. The time of light propagation from point *a* to point *b*, $\Delta t_{ab}$, must be equal to *L/(c–v)*, where *L* – the distance between points *a* and *b*, while the light propagation time in the reverse direction, $\Delta t_{ba}$, is *L/(c+v)*. It was assumed that the total time of light propagation from point *a* to point *b* and back to point *a*, $\Delta t_{aba}$, is determined by the formula

$$\Delta t_{aba} = \frac{2L}{c(1 - v^2/c^2)}. \qquad (I.1)$$

It must be different from the velocity in the ether, *c*, according to the concepts of this time, and the speed of light propagation in the laboratory in the direction perpendicular to the laboratory's longitudinal axis, which is aimed along the line of travel. In this instance, the lateral speeds of light propagation from one of the side walls to another and in the reverse direction, $c_{lat}$, must be equal to one another, and are linked to the *c* and *v* velocities by the formula $c_{lat} = c\sqrt{1 - (v/c)^2}$. Consequently, the time required for light to traverse a distance of *L* from point *a* to point *d* and back to point *a* in the lateral direction, $\Delta t_{ada}$, should be linked to distance by the correlation

$$\Delta t_{ada} = \frac{2L}{c\sqrt{1 - (v/c)^2}}. \qquad (I.2)$$

It seemed obvious that, as is clear from formulas (I.1) and (I.2), the ratio of $\Delta t_{aba}$ to $\Delta t_{ada}$ when the distance of points *b* and *d* from point *a* is identical, must be equal to $1/\sqrt{1 - (v/c)^2}$. At the time, the technical capability existed for the indirect experimental determination of this ratio, and Michelson and Morley did it. The experiment indirectly demonstrated the equality of the $\Delta t_{aba}$ and $\Delta t_{ada}$ time values.

At first, this result seemed strange and inexplicable to the physicists of the late 19$^{th}$ century. However, FitzGerald and Lorentz quite quickly found an explanation for the result obtained. According to the explanation of FitzGerald and Lorentz, when moving through the ether, a body interacts with the latter and is shortened by $1/\sqrt{1-(v/c)^2}$ times. Due to this shortening, the time of light propagation back and forth in the longitudinal direction, $\Delta t_{aba}$, proves to be $1/\sqrt{1-(v/c)^2}$ times less than anticipated, and consequently, is exactly equal to the light propagation time back and forth in the lateral direction.

The possibility of detecting an ethereal wind by means of measuring the $\Delta t_{ada}$ time, and based on it, the lateral velocity, $c_{lat}$, equal to $2L/\Delta t_{ada}$, would seem to follow from formula (I.2), thereby doing away with the effect of longitudinal contraction on the measurement result. Michelson and Morley were unable to perform such a measurement for technical reasons; however, it is clear to us now that the $\Delta t_{ada}$ time measured by a moving clock would not be increased by $1/\sqrt{1-(v/c)^2}$ times, as formula (I.2) indicates, due to the slowing of the rate of the moving clock not taken into account therein. This slowing offsets the anticipated increase in the $\Delta t_{ada}$ time. The $c_{lat}$ velocities in the ether and in the moving laboratory would be in agreement. An experiment that Kennedy and Thorndike conducted using an asymmetrical interferometer confirms this effect.

In popular and educational books on the theory of relativity, it is frequently claimed that Michelson and Morley experimentally proved the fallacy of the ethereal concept. This is an untrue statement. The outcome of the Michelson-Morley experiment demonstrated the impossibility of detecting ether using an experiment of this type, but not the absence of ether. One must not lose sight of the fact that the results of the Michelson and Morley experiment did not prevent the notions about the ether from being retained for a good two decades after they obtained their negative measurement result. Michelson himself and Lorentz shared these notions. Indeed, FitzGerald and Lorentz explained this result by way of the longitudinal shortening of objects moving through the ether, i.e., they explained it within the framework of the ethereal world view. And so it was that in the waning years of his life, in 1952, Einstein wrote in the article "Relativity and the Problem of Space": Concerning the experiment of Michelson and Morley, H. A. Lorentz showed that the result obtained at least does not contradict the theory of an ether at rest" [3].

In this regard, the remark of a proponent and popularizer of the theory of relativity, M. von Laue, should also be clear, who wrote: "…it was experimentally impossible to make a choice between this theory (the Lorentz theory) and Einstein's theory of relativity, and if the Lorentz theory nonetheless took a back seat – even though it still has proponents among physicists – this then undoubtedly occurred due to reasons of a philosophical nature" [4].

How should M. von Laue's remark that it is experimentally impossible to make a choice between the Lorentz theory and the Einstein theory be taken? Indeed, according to Lorentz, reference systems at rest in the ether and inertially absolutely moving through the ether are not physically equal. Is this circumstance really in agreement with the fact of the equality of inertial systems in Einstein's STR?

It is.

The equality of inertial systems in the STR is expressed by way of the invariance of the mathematical notation of the laws of nature in these systems, but this form of equality historically migrated to Einstein's STR from the Lorentz theory. Indeed, the transformations that ensure the immutable form of notation of these same Maxwell equations in different, physically unequal inertial reference systems appear in Lorentz's work as a consequence of the requirement for such immutability. While Einstein tied the immutability of the form of notation of the physical laws of nature in different reference system to physical equality, Lorentz demonstrated that this requirement can be met even in physically unequal systems absolutely at rest and absolutely in motion. I.e., Lorentz showed that physically unequal inertial reference systems can be transformed into mathematically equal ones by imposing the requirement of invariance on them.

Is this possible or not? The fact that it is possible is revealed on the pages of this article. Two groups of barges are examined in the article, one of which is at rest on a water surface, while the other is in motion relative thereto at a velocity of $v$. Primal processes occur on the barges, information on the progress of which is transferred from barge to barge by auxiliary boats that have a constant velocity of $V$ relative to the water. As concerns the barge group in motion, the velocity of the auxiliary boats in their direction of movement equals $V - v$, while in the direction opposite that of their movement, it is $V + v$. The processes occurring on the barges make it possible to fully simulate STR effects, including Lorentz contraction, time dilation, relativistic Doppler effects – both longitudinal and lateral – and the relativistic law of the addition of velocities. An analogy of the STR simulation model is achieved by assuming the equality of the velocities of the auxiliary boats relative to the barge group in motion in different directions. The possibility of presenting the simulated effects using a simulated model of space-time provides food for thought concerning the relationship of formal space-time to the real world. It is possible to arbitrarily interpret the material contained in the article based on an examination of the movement of the barges on the water as it relates to Einstein's STR, but a fact is still a fact – in the physically unequal and physically asymmetrical coordinate systems linked to the barge groups at rest on the water surface and in motion on the water, the simulation of the full symmetry of the processes being observed and the equality of the systems with reference to the formal description of these processes is possible.

**1. Subjects and Essence of the Simulation**

In the body of the text, we will not go into the details, which can be examined separately from the description of the essence of the matter without detriment to the comprehension of the material.

Let us mentally observe "from the outside" the behavior of objects that, being slow-moving, nonetheless act in accordance with laws similar to those of the special theory of relativity. Two groups of barges distributed over the surface of a boundless flat-bottomed water body – group $R$ and group $R'$ – will serve as the subjects of our mental observation. Barge group $R$ is at rest on the surface, while the group $R'$ barges are in motion on this surface at the same velocity, $v$, and in the same direction. In examining the geometric dimensions of sections of the barge groups, we will regard the barges themselves as point objects, the dimensions of which are negligibly small as compared to those of the barge group sections.

We will presume that the water in the water body is slack. The depth of the water body is universally identical and equals $h$. On each barge, high-speed underwater shuttles that have a velocity of $V$ (relative to the water) deliver sand from the bottom of the water body. We will call the watercraft that, under our assumption, have an identical velocity of $V$ not attainable by the other watercraft high-speed. One high-speed cargo shuttle is assigned to each barge and a specific amount of sand is placed into each such shuttle, which we will call one shuttle of sand. Each shuttle continually makes shuttle trips from a barge to the bottom and back over the shortest possible route, moving along a plumb line that runs from a given barge to the bottom. The times required for a shuttle to pick up the sand from the bottom and to offload it from the shuttle onto a barge are negligibly small as compared to the shuttle movement time from the barge to the bottom and back. Thus, the time required to deliver the sand to a barge at rest, $\Delta t$, is given by the formula

$$\Delta t(1) = 2h/V \qquad (1)$$

for one shuttle of sand, and by the formula

$$\Delta t(k) = 2kh/V \qquad (2)$$

for *k* shuttles of sand. By the time denoted with the letter *t*, we mean the time read by our clock – the clock of the outside observers. The number one and the letter *k* in parentheses associated with the Δ*t* symbol in formulas (1) and (2) show that reference is being made the Δ*t* time interval needed in order to deliver one or *k* shuttles of sand, respectively.

The plumb movement velocity (the submersion and surfacing speed), $V_Z$, of a shuttle delivering sand to a barge moving at a velocity of *v* (*v* < *V*) is slower than the *V* velocity and is given, as is clear from Fig. 1, by the formula

$$V_Z = V\sqrt{1-(v/V)^2}. \qquad (3)$$

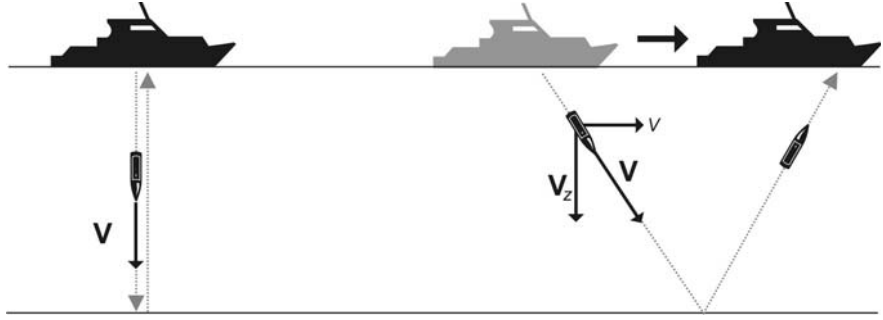

*Fig. 1. Barge r (on the left) is at rest on the water surface. A shuttle is moving from the barge to the bottom and back at a velocity of V. Barge r' (on the right) is in motion on the water body surface at a velocity of v. The speed of movement of the shuttle equals V, the shuttle's horizontal velocity component equals v, and its vertical component, $V_Z$, equals $V\sqrt{1-(v/V)^2}$.*

Thus, a time of $\Delta t(1') = 2h/V_Z$ is needed in order to deliver one shuttle of sand to a barge drifting at a velocity of *v*, or with allowance for formula (3)

$$\Delta t(1') = \frac{2h}{V\sqrt{1-(v/V)^2}}, \qquad (4)$$

The prime near the number one means that reference is being made to the delivery of sand to a barge in motion (not at rest).

Delivering *k'* shuttles of sand to a barge proceeding at a velocity of *v* requires a time of:

$$\Delta t(k') = \frac{2k'h}{V\sqrt{1-(v/V)^2}}. \qquad (5)$$

The letter *k'* in parenthesis in the notation Δ*t*(*k'*) indicates that reference is being made to the Δ*t* time interval needed in order to deliver *k'* shuttles of sand. As in the case of the primed number one, the presence of a prime near the letter *k* suggests the delivery of sand to a barge in motion (not at rest).

Among the barges, if there is a barge *r'* that is in motion on the water surface at a velocity of *v* and a barge *r* that is at rest on the water, the rate of increase in the quantity of sand on barges *r'* and *r* will be different. It follows from formulas (2) and (5) that at *k'* = *k*

$$\Delta t(k') = \Delta t(k) \big/ \sqrt{1 - (v/V)^2} \ , \qquad (6)$$

while at $\Delta t(k') = \Delta t(k)$

$$k' = k\sqrt{1 - (v/V)^2} \ , \qquad (7)$$

i.e., the times required for delivering an identical quantity of sand to a certain barge *r′* in motion at a velocity of *v* and to a certain barge *r* at rest differ from one another by $1\big/\sqrt{1-(v/V)^2}$ times. The quantities of sand delivered to barge *r* at rest and to barge *r′* in motion when the time for its delivery is identical also differ from one another by as many times.

Let's say that the barges have a roughly identical maximum load-carrying capacity, and assuming that the quantity of sand onboard is close to critical, they sink and cease to exist, whereupon new barges appear to replace them, on which there is no sand at the time of their appearance ("birth"). If the critical quantity of sand for the barges in motion equals an average of *K′*, then according to formula (5), the average "lifetime" of the barges, $\overline{\Delta t}(K') = \dfrac{2K'h}{V\sqrt{1-(v/V)^2}}$, from the moment of their "birth", which consists of the commencement of the loading of sand onto them, till their "death", is dependent upon their speeds of movement. If it is presumed that the critical quantity of sand for the barges in motion and those at rest is identical, i.e., if *K′* = *K*, then the barges in motion at a velocity of *v* will live an average of $1\big/\sqrt{1-(v/V)^2}$ times longer than the barges at rest.

We are talking about the average lifetime of an individual barge, assuming that the lifetime of an actual barge might differ somewhat from the average lifetime due to the approximate nature of the critical quantity of sand, or due to unforeseen circumstances (for example, emergency situations). For this reason, we assume that even among barges residing in an identical kinematic state (for example, in a state of rest), a host of "mixed-age" barges with different quantities of sand onboard is present on the water body's surface.

The dependence of the average lifetime of the barges upon speed of movement, as well as some of their other properties, is a convenient feature that makes it possible to simulate the special theory of relativity using elementary methods of classical mechanics. Simulation yields Lorentz transformations, the relativistic addition of velocities, the effects of relativistic time dilation and the contraction of the longitudinal dimensions of moving objects, the relativity of simultaneity, and the relativistic Doppler effect, including the transverse effect. In general, simulation yields all the effects of the special theory of relativity and is easily presented in the form of a model of the four-dimensional space-time world. Moreover, simulation affords the opportunity of clearly seeing the causes of the relativistic effects and paradoxes of the special theory of relativity, such as the twin paradox for instance, and makes it possible to answer not only the question of "How?", but also the question of "Why?".

At first glance, it is impossible to simulate not only the special theory of relativity, but even the time dilation it describes based on the slowing of the rate of sand loading onto the barges during their movement, if only for the reason that the special theory of relativity is a relativistic theory. The slowing of process time on one of two identical bodies, *a* and *b*, in motion relative to one another occurs therein relatively and symmetrically. The relativity and symmetry are manifested by way of the fact that if the processes in body *b* are slowed relative to the reference system associated with body *a*, the processes in body *a* are then also exactly slowed relative to the reference system associated with body *b*. In our case, however, the increase in the quantity of sand on the barge in motion, *r′*, occurs more slowly than the increase in the quantity of sand on the barge at rest, *r*, absolutely and asymmetrically. The absoluteness and asymmetry are manifested by way of the fact that if the quantity of sand on the barge in motion increases more slowly than the quantity of sand on the barge at rest, this fact is not dependent upon the barge from which these processes are observed. This would seem to make the model involving barges obviously unacceptable for simulating the special theory of relativity.

But we will not jump to conclusions! We are after all talking about a model, and the model's objective consists not of its complete oneness with the object being modeled, but rather the formal equivalence of the phenomena it describes under specific conditions to the phenomena that are observed within the object being modeled.

**2. Hardware Components on the Barges. Primary Objective of the Hardware Components**

We will assume that we, as outside observers, have all the essential tools for observing the barges, as well as the devices for measuring physical quantities. We will also assume that there are no observers on the barges – hardware measuring devices have been placed on them that possess limited capabilities. The makeup of the hardware components on the barges does not include long rulers or rods that connect the individual barges, and all communications and interactions between the barges are accomplished using speedboats that travel between them. Our conventional clocks that consistently measure time on the barges at rest and in motion are not present on the barges – all processes on the barges occur in simulated time, which we will talk about further on in this text.

There are no instruments on the barges that make contact with the water, which would make it possible to record an instance of barge motion or rest relative to the water. There are also no optical instruments and radio devices on the barges that are capable of almost instantaneously transmitting information from barge to barge. The transfer of information from barge to barge on physical media is only accomplished directly (when the barges are located in direct proximity to one another) or using speedboats (that ply the water surface at a velocity of $V$), which carry the physical information media from one barge to another. Only the information contained on the physical media is regarded as documentary (documented) information.

The primary objective of the hardware components, having information at their disposal concerning the essence of the environment and concerning the fact that the rate of sand loading onto the barges at rest in the environment is higher than the rate of sand loading onto the barges in motion, consists of determining the means for exchanging documented information on the quantity of sand on the different barges, the barges that the sand reaches more quickly, and accordingly which of them are at rest. As strange as it sounds, it is impossible to solve this seemingly simple problem given the limitations cited.

**3. Symmetry of Recording the Processes of Sand Delivery to the Barges from a Barge at Rest and a Barge in Motion**

If the barges are inertial and are either not encountered at one point at all or are only encountered once, changes in the quantity of sand on the barges can only be compared using speedboats (traveling at a velocity of $V$), on which documents concerning the quantity of sand are delivered from one barge to another. For example, after meeting a barge at rest, $r$, when information concerning the quantity of sand on the barges is directly transferred from craft to craft, if barge $r'$ continues to move in a straight line, while barge $r$ remains motionless (barges $r$ and $r'$ are inertial), barges $r$ and $r'$ then cannot meet a second time. Thus, it is only possible to again obtain information concerning the quantity of sand on the other barge some time after a meeting if a boat is used. In this instance, a document indicating the quantity of sand on the other barge reaches this barge when the quantity of sand on the other barge is already not the same as it was at the time that the boat was dispatched from it. Only if the $V$ velocity is much greater than the $v$ velocity, i.e., if $V \gg v$, can the information be regarded as having been transferred instantaneously. In this case, the changes in the quantities of sand that occur between the time that the barges meet and the time that the document is transferred are directly compared. But, here, the $1\!\left/\!\sqrt{1-(v/V)^2}\right.$ coefficient is almost equal to unity and the comparison results are symmetrical by virtue of the sameness of the sand increase rates on the barge at rest and the barge in motion. However, if the $V$ and $v$ velocities are commensurate, the $1\!\left/\!\sqrt{1-(v/V)^2}\right.$ coefficient is then sufficiently great and the difference in the sand loading rates is great, but the effect of the delay in information transfer by boat is also great. The consequence of this effect is such that when information is transferred by boat and when conventional, identically running clocks are not present on the barges at rest and in motion, then it is only possible to obtain symmetrical results, which do not permit the detection of the difference in the rates of increase of the quantity of sand on the barge at rest, $r$, and the barge inertially in motion, $r'$.

Yet another method is suggested for comparing the rates at which the sand reaches the barges.

If the number of boats for transferring information is unlimited, the hardware components on each barge can then continuously send information concerning the arrival of the next, preagreed – signal – portion of sand at their barge by dispatching a boat to the other barge each time that carries information on said arrival. This makes it possible to compare the frequency of arrival of signal portions of sand at a given barge to the frequency of the arrival of boats at it from the other barge.

According to the classical Doppler effect formula, the frequency of the arrival of the boats dispatched from barge $r'$ at barge $r$, $f_{r\leftarrow r'}$, and the frequency of the arrival of the boats dispatched from barge $r$ at barge $r'$, $f_{r'\leftarrow r}$, will be dependent upon the velocity, $v$, of barge $r'$ relative to the water, as well as upon which of the barges plays the role of the transmitter and which plays the role of the receiver of the information. As barge $r'$ moves away from barge $r$, the $f_{r\leftarrow r'}$ and $f_{r'\leftarrow r}$ frequencies are given by the formulas

$$f_{r\leftarrow r'} = f_{r'}/(1+v/V),$$
$$f_{r'\leftarrow r} = f_r(1-v/V),$$

where $f_r$ and $f_{r'}$ – the frequencies of the arrival of signal quantities of sand at barges $r$ and $r'$, respectively, and the frequencies of the dispatch of boats from barges and $r$ and $r'$, respectively, which are numerically equal to them.

According to these formulas, when the $v$ velocity of barge $r'$ approaches the $V$ velocity, the frequency of boat arrival at the barge at rest, $r$, $f_{r\leftarrow r'}$, approaches $f_{r'}/2$, while the frequency of boat arrival at the barge in motion, $r'$, $f_{r'\leftarrow r}$, approaches zero at a finite frequency of $f_r$.

As barge $r'$ approaches barge $r$, the $f_{r\leftarrow r'}$ and $f_{r'\leftarrow r}$ frequencies are then given by the formulas:

$$f_{r\leftarrow r'} = f_{r'}/(1-v/V),$$
$$f_{r'\leftarrow r} = f_r(1+v/V).$$

When the $v$ velocity of barge $r'$ approaches the $V$ velocity and when the $f_{r'}$ value is different from zero, the frequency of boat arrival at the barge at rest, $r$, $f_{r\leftarrow r'}$, approaches infinity, while the frequency of boat arrival at barge $r'$, $f_{r'\leftarrow r}$, approaches $2f_r$.

If the $f_r$ and $f_{r'}$ frequencies are to equal one another, the $f_{r\leftarrow r'}$ and $f_{r'\leftarrow r}$ frequencies would then be different at $v \neq 0$. The $f_{r\leftarrow r'}/f_r$ and $f_{r'\leftarrow r}/f_{r'}$ ratios of the frequencies of boat arrival at a barge to the frequencies of signal quantities of sand reaching them would also be different. However, in our case, the $f_r$ and $f_{r'}$ frequencies are linked to one another by the correlation $f_{r'} = f_r \sqrt{1-(v/V)^2}$. For this reason, the following equalities can be written using the previous formulas

$$f_{r \leftarrow r'} = f_r \sqrt{1-v/V}/\sqrt{1+v/V},$$
$$f_{r' \leftarrow r} = f_{r'} \sqrt{1-v/V}/\sqrt{1+v/V}$$

for the case of barge $r'$ moving away from $r$, as can the equalities

$$f_{r \leftarrow r'} = f_r \sqrt{1+v/V}/\sqrt{1-v/V},$$
$$f_{r' \leftarrow r} = f_{r'} \sqrt{1+v/V}/\sqrt{1-v/V}$$

for the case of their convergence.

The $f_{r \leftarrow r'}/f_r$ and $f_{r' \leftarrow r}/f_{r'}$ ratios of the frequencies of boat arrival at each barge to the frequencies of signal quantities of sand reaching it for the barge at rest on the water and for the barge in motion are identical and equal $\sqrt{1-v/V}/\sqrt{1+v/V}$ as barge $r'$ moves away from barge $r$, or $\sqrt{1+v/V}/\sqrt{1-v/V}$ as barge $r'$ approaches barge $r$.

If even one of the barges violates the condition of inertness, it is then possible to ensure that barges $r$ and $r'$, having met once, will meet again some time later. For example, if some time after the barges meet, barge $r'$, having changed its direction of movement, returns to the first meeting point at the same velocity (barge $r'$ becomes noninertial at the time that it turns), it is then possible to ensure that the barges meet again and to make a second direct comparison of the quantities of sand on the barges. The data acquired will make it possible to determine the increase in sand on the barges during the time between the two barge meetings. In this instance, it becomes clear that the increase in the quantity of sand on barge $r'$ during this period is Γ times smaller than the increase in the quantity of sand on barge $r$. It is not difficult for us, as outsiders, to calculate the Γ coefficient. It equals $1/\sqrt{1-(v/V)^2}$.

However, this experiment does not afford the opportunity, based on the results obtained and the information concerning the slowness of the processes of sand loading onto the barge in motion, to achieve the primary objective and to reach the conclusion that barge $r$ is at rest in the environment. The fact is that, after barges $r$ and $r'$ meet, if barge $r'$ continues moving, while barge $r$ remains at rest for a certain, let's say, arbitrarily long amount of time, then leaving the train of barges at rest, picks up the necessary speed and catches up to barge $r'$, the increase in the sand on barge $r$ would be smaller when barges $r$ and $r'$ meet. Here, this decrease would be such that it might be interpreted as proof that barge $r'$ is at rest in the environment, while barge $r$ is in motion back and forth. Based on the results of the experiments involving inertial barges, it is only possible to establish the fact that the slowing of the increase in the quantity of sand on one of the two barges during the period of time between the two meetings can be tied to its inertness, during which, formally, this slowing is qualitatively and quantitatively similar to the slowing of the aging of one of two twins in the well-known paradox.

**4. Simulation of Time on Individual Barges. Conventional and Simulated Times**

Let's say that counters of the quantities of sand reaching the barges and identical clocks with identical faces have been placed on all the barges. The clocks are triggered by the sand quantity counters in such a way that the clock on each barge "ticks" at a frequency proportional to that of shuttle movement. The hands of the clocks on the barges traveling at a velocity of $v$ move $1/\sqrt{1-(v/V)^2}$ times more slowly than the hands of the clocks on the barges at rest. We will presume that the speed of movement of the clock hands on the barges at rest equals the speed of movement of the hands of our clock – the clock of the outside observers – i.e., time, $t$, including its numeric values, also flows identically for us on the barges at rest. We will denote the reading of a clock on a certain individual barge at rest, $r$, using the symbol $t_r$. Here, we will assume that the readings, $t_r$, $t_a$, $t_b$, and $t_c$, of the clocks on different barges at rest, $r$, $a$, $b$, and $c$, can differ from one another at the same moment in time up to the point that these readings are no longer synchronized. We do not view this difference in readings when the rate of their replacement is identical as a difference in the passage of time, just as we do not presume that our terrestrial time, $t$, passes differently in London and in Moscow, even though the time (the clock readings) in London $t_L$, and in Moscow, $t_M$, differ from one another at each moment in time. By the sameness of the passage of time, we mean not the coincidence of the $t_L$ and $t_M$ clock readings, which does not occur, but rather than equality of the $\Delta t_L$ and $\Delta t_M$ time intervals in these cities between two specific world events.

Unlike our time, $t$, we will call the time given by the readings of the clocks on the barges in motion at a velocity of $v$ the simulated time of the barges in motion and we will denote it using the letter $t'$ (with a prime). Concerning the simulated time on the barges in motion, $t'$, we will say that, based on our observation results, it passes more slowly that the time on the barges at rest, $t$. We will denote the $t'$ clock reading on a certain individual barge in motion, $r'$, using the symbol $t'_{r'}$ (with primes), assuming that, as in the case of the barges at rest, the clock readings on the other barges in motion may differ from one another and from the $t'_{r'}$ reading up until the moment of their synchronization.

We will presume that the signals from the sand quantity counters proceed in the form of sync pulses not only to the clocks, but also to all the hardware components located on the barges without exception, thereby assigning their operating rate and ensuring hardware component performance on the simulated time scale. The operating rate (the response time) of the hardware components on the barges in motion will then be slower than the operating rate of the hardware components on the barges at rest by $1/\sqrt{1-(v/V)^2}$ times.

Let's say that the time intervals $\Delta t(\Delta n_{time})$ and $\Delta t'(\Delta n'_{time})$ are used as the units of measurement of the time on the barges at rest and in motion, respectively, over the course of which the sand quantity counter readings increase by $\Delta n_{time}$ and $\Delta n'_{time}$, respectively. so that, being common and standard, $\Delta n_{time} = \Delta n'_{time}$. In this instance, the $\Delta t$ time interval and the $\Delta t'$ simulated time interval for any pair of barges in motion relative to one another between two identical events that we record are linked by the correlations

$$\Delta t' = \Delta t\sqrt{1-(v/V)^2} \text{ and } \Delta t = \Delta t'/\sqrt{1-(v/V)^2}. \qquad (8)$$

If, for any reason, the $\Delta t(\Delta n_{time})$ time measurement unit on the barges at rest has the same denomination as the $\Delta t'(\Delta n'_{time})$ simulated time unit on the barges in motion, for example, if both these units are called a minute, then according to our observations, it is possible to say that a minute of simulated time on the barges in motion lasts $1/\sqrt{1-(v/V)^2}$ times longer than a minute of time on the barges at rest.

If the number of operations they perform dictates the service life of the individual devices that make up the hardware components, then according to our observation results, the service life (the "lifetime") of the devices on the barges in motion will be the same $1/\sqrt{1-(v/V)^2}$ times greater than on the barges at rest. I.e., the "life" of the observation devices located on the barges in motion will pass more slowly than the "life" of the similar devices on the barges at rest. We note that, due to the synchronism of their own operations with those of the auxiliary devices and with the movement of the shuttles, the hardware components themselves, which may also include automatic devices, on the barges at rest and in motion "perceive" the change in the quantity of sand over the course of their lives and all the processes that occur synchronously with the operation of the sand counters on their barges as entirely identical. It is impossible for any of the instruments that are a part of the hardware components on each of the barges in motion to detect the slowness of the operating rates on a barge in motion. Due to the synchronism of the operation of these instruments with that of the other instruments and devices, all the processes and operations on the barges in motion in simulated time also occur exactly like the similar processes on the barges at rest in conventional time.

It is not difficult to see that, as noted in the previous section, if it is impossible to detect the difference in the rates at which the sand reaches the barges by means of a direct comparison, then due to the proportionality of the counter readings to the readings of the clocks on the barges, it is also impossible to detect the difference in the passage of conventional time on the barges at rest and simulated time on the barges in motion using speedboats to transfer the clock reading information. Moreover, it is impossible to detect the difference in the passage of time by simulating the Doppler effect within the aquatic environment in which the barges are located. For example, if boats are dispatched from barge $r$ to barge $r'$ with a frequency of $f_r$ and boats are dispatched from barge $r'$ to barge $r$ with a frequency of $f'_{r'}$ (in simulated time), which is numerically equal to the $f_r$ frequency, it is then not difficult to show that as barge $r'$ moves away from barge $r$, the $f'_{r' \leftarrow r}$ and $f_{r \leftarrow r'}$ frequencies of boat arrival at barges $r'$ and $r$, respectively, are identical and are given by the symmetrical formulas $f'_{r' \leftarrow r} = f_r \sqrt{1 - v/V} / \sqrt{1 + v/V}$ and $f_{r \leftarrow r'} = f'_{r'} \sqrt{1 - v/V} / \sqrt{1 + v/V}$, which, being similar to the relativistic Doppler effect formulas, do not make it possible to detect the difference in the clock rate on barges $r$ and $r'$.

As the barges converge, the $f'_{r' \leftarrow r}$ and $f_{r \leftarrow r'}$ frequencies of boat arrival at barges $r'$ and $r$, respectively, are identical and are given by the symmetrical formulas $f'_{r' \leftarrow r} = f_r \sqrt{1 + v/V} / \sqrt{1 - v/V}$ and $f_{r \leftarrow r'} = f'_{r'} \sqrt{1 + v/V} / \sqrt{1 - v/V}$.

If barge $r'$ moves along a line from which barge $r$ is located a certain distance way, the observation of a "boat-type relativistic" transverse Doppler effect is then possible, during which the equalities $f'_{r' \leftarrow r} / f_r = f_{r \leftarrow r'} / f'_{r'} = \sqrt{1 - (v/V)^2}$ hold true.

**5. Simulation of the Lorentz Contraction of the Distance Between Moving Elements**

We will now tie the $\Sigma$ and $\Sigma'$ Cartesian coordinate systems with mutually parallel axes of $X$ and $X'$, $Y$ and $Y'$, and $Z$ and $Z'$ to the barge groups at rest, $R$, and in motion, $R'$, respectively. We will direct the system $Z$ and $Z'$ axes perpendicular to the water surface and the $X$ and $X'$ axes in the direction of movement of the $R'$ group, while we will lead the $Y$ and $Y'$ axes perpendicular to the $X$, $X'$ and $Z$, $Z'$ axes, which is acceptable to do in rectangular Cartesian coordinate systems. We will presume that the hardware components of each of the barges are capable of independently measuring the distance from a given barge to the individual points of the coordinate system tied to it without the involvement of the hardware components of the other barges and without the synchronization of the clocks on the different barges. According to our proposition, the distance is measured remotely without using long rulers or tape measures in the following manner.

Let's assume that in the $\Sigma$ coordinate system, the hardware components on barge $r$, which is located at the origin of coordinates, $O$, determine the distance from the origin of coordinates, $O$, to a certain point, $a$, at an arbitrary location in the $\Sigma$ system using a method that we will call pseudoradar. The essence of the pseudoradar method is as follows. A speedboat is dispatched from barge $r$ at point $O$, which goes to point $a$ using the shortest possible route, and once there, goes back to point O. The boat's travel time back and forth, $\Delta t_{OaO}$, is determined at the moments of its departure and return using the readings of the clock on barge r, then the distance between points O and a is calculated using the formula $\frac{1}{2}\widetilde{V}\Delta t_{OaO}$, where $\widetilde{V}$ – the average velocity of the boat en route from point O to point a and back. Since velocity $\widetilde{V}$ is not determined based on distance and time in the Σ system, but rather distance is determined based on velocity $\widetilde{V}$ and time, a velocity value that is standard and unified for all the Σ system barges can either be assigned by the hardware components of the Σ system barges or taken from the outside (for example, from us). It is clear that in the Σ system, the average velocity en route back and forth, $\widetilde{V}$, equals the V velocity. We will denote the distance between point O and point a measured in this manner by the hardware components on barge r using the symbol $l(\frac{1}{2}\Delta t_{OaO})$. We will denote this same distance, but measured or calculated by us in any available way, using the symbol $l_{Oa}$. The $l(\frac{1}{2}\Delta t_{OaO})$ distance and the $l_{Oa}$ distance are one and the same, while the $l(\frac{1}{2}\Delta t_{OaO})$ and $l_{Oa}$ quantities only differ from one another by the method and the location in which they are determined.

If the barges change their position due to certain external causes (for example, because of the wind), the hardware components can then maintain the distance between the barges in a given group using the speedboats. To this end, speedboats are regularly dispatched from each barge to the neighboring barges and back again. Using their clocks, the barge hardware components measure the boat movement time to the neighboring barges and back, then approach or move away from the neighboring barges as necessary in order to maintain these times and the immutability of the pseudoradar distances.

We will presume that in the Σ' coordinate system, the hardware components of a certain barge r', located at the origin of coordinates, O', determines the distance from the origin of coordinates, O', to point a' in the Σ' system exactly as this is done in the Σ system. Let's say that a quantity equal to the product of $\frac{1}{2}\widetilde{V}'\Delta t'_{O'a'O'}$ is by definition regarded as the distance measured in this manner in the case at hand, $l'(\frac{1}{2}\Delta t'_{O'a'O'})$. Here, $\widetilde{V}'$ is simulated, i.e., the average velocity of a boat en route from point O' to point a' and back, expressed through simulated time, t', is by definition numerically equal to the $\widetilde{V}$ velocity and consequently the V velocity, while $\Delta t'_{O'a'O'}$ is the simulated time of movement of the boat en route from point O' to point a' and back. Unlike the conventional average velocity, $\widetilde{V}$, we denote the simulated average velocity using the symbol $\widetilde{V}'$ (with a prime!). We will call the pseudoradar distance determined in this manner by the hardware components of barge r', $l'(\frac{1}{2}\Delta t'_{O'a'O'})$, the simulated distance, $l'(\frac{1}{2}\Delta t'_{O'a'O'})$. The simulated distance has two features that are useful to us. First, by definition, it ensures the equality of the simulated average (en route back and forth) velocity, $\widetilde{V}'$, and the V velocity. And second, based on our observation results, it equals the conventional distance in a direction that is transverse relative to Σ' system movement, but is $1/\sqrt{1-(v/V)^2}$ times greater than the conventional distance in the direction of Σ' system movement. I.e., the conventional distances that we measure between the elements of the Σ' movement system in the direction of movement are $1/\sqrt{1-(v/V)^2}$ times shorter than the distances between the subject elements in this system.

In point of fact, if a speedboat moves along the Y' axis between point O' (the origin of coordinates) and a point with a coordinate of y' that lies at the Y' axis – we will call this point the y' point, the boat's V velocity component that we record from the outside (the speed of movement to the boat along a segment of a straight line that connects points O' and y'), $V_Y$, is then given by the formula

$$V_Y = V\sqrt{1-(v/V)^2}. \qquad (9)$$

According to our calculations, the conventional distance between points $O'$ and $y'$ in the $\Sigma'$ movement system that we record, $l_{O'y'}$, can be represented by the formula $l_{O'y'} = \frac{1}{2} V_Y \Delta t_{O'y'O'}$, where $\Delta t_{O'y'O'}$ – the boat's conventional time of movement (that we measure using our clock) from point $O'$ to point $y'$ and back, or with allowance for formula (9)

$$l_{O'y'} = \tfrac{1}{2} V \sqrt{1-(v/V)^2}\, \Delta t_{O'y'O'}, \qquad (10)$$

while according to barge $r'$ hardware component data, the simulated distance between points $O'$ and $y'$, $l'(\tfrac{1}{2}\Delta t'_{O'y'O'})$, is expressed by the formula

$$l'(\tfrac{1}{2}\Delta t'_{O'y'O'}) = \tfrac{1}{2} \widetilde{V}' \Delta t'_{O'y'O'}. \qquad (11)$$

where $\Delta t'_{O'y'O'}$ is the boat's time of movement from point $O'$ to point $y'$ and back, simulated on the barge at point $O'$.

By analogy with the left-hand equality of correlation (8), since

$$\Delta t'_{O'y'O'} = \Delta t_{O'y'O'} \sqrt{1-(v/V)^2},$$

while $\widetilde{V}'$ by definition equals $V$, it then follows from formulas (10) and (11) that

$$l_{O'y'} = l'(\tfrac{1}{2}\Delta t'_{O'y'O'}), \qquad (12)$$

which suggests the equality of the numeric values of the conventional distance, $l_{O'y'}$, and the simulated distance, $l'(\tfrac{1}{2}\Delta t'_{O'y'O'})$, and in generalized terms, implies the equality of the numeric values of the lateral dimensions.

We will now examine the movement of a speedboat between point $O'$ (the origin of coordinates) and a point with a coordinate of $x'$ that lies at the $X'$ axis – we will call it point $x'$. According to our data, the boat's speeds of movement relative to these points in opposite directions equal $V - v$ and $V + v$. At the conventional distance between points $O'$ and $x'$, which equals $l_{O'x'}$, the boat's time of movement from point $O'$ to point $x'$ and back, $\Delta t_{O'x'O'}$, equals $l_{O'x'}/(V-v) + l_{O'x'}/(V+v)$, i.e.,

$$\Delta t_{O'x'O'} = \frac{2 l_{O'x'}}{V(1-v^2/V^2)}. \qquad (13)$$

According to our data, the boats average speed of movement along the $X'$ axis relative to points $O'$ and $x'$ en route back and forth, $\widetilde{V}_{X'}$, equals $2 l_{O'x'}/\Delta t_{O'x'O'}$, or with allowance for the previous equality,

$$\widetilde{V}_{X'} = V(1-v^2/V^2). \qquad (14)$$

According to our calculations, the conventional distance between points $O'$ and $x'$ in the $\Sigma'$ movement system, $l_{O'x'}$, can be expressed by the formula

$$l_{O'x'} = \tfrac{1}{2} \widetilde{V}_{X'} \Delta t_{O'x'O'}, \qquad (15)$$

while according to the calculations of the hardware components on barge $r'$, the simulated distance between them, $l'(\tfrac{1}{2}\Delta t'_{O'x'O'})$, is expressed by the equality

$$l'(\tfrac{1}{2}\Delta t'_{O'x'O'}) = \tfrac{1}{2}\widetilde{V}'_{X'}\Delta t'_{O'x'O'}. \tag{16}$$

By analogy with the right-hand equality of correlation (8), since $\Delta t_{O'x'O'} = \Delta t'_{O'x'O'}\big/\sqrt{1-(v/V)^2}$, then taking formula (14) into account, it follows from formulas (15) and (16) that

$$l_{O'x'} = \sqrt{1-(v/V)^2}\, l'(\tfrac{1}{2}\Delta t'_{O'x'O'}) \ \ \text{or}\ \ l'(\tfrac{1}{2}\Delta t'_{O'x'O'}) = l_{O'x'}\big/\sqrt{1-(v/V)^2}\ . \tag{17}$$

According to formula (17), the numeric value of the distance that we measure between points $O'$ and $x'$, $l_{O'x'}$, is $1\big/\sqrt{1-(v/V)^2}$ times smaller than the numeric value of the distance measured in the $\Sigma'$ coordinate system, $l'(\tfrac{1}{2}\Delta t'_{O'x'O'})$. In generalized terms, this suggests the contraction of sections of the group in motion, $R'$, by $1\big/\sqrt{1-(v/V)^2}$ times.

Strictly speaking, formulas (12) and (17) should be written in the form

$$\{l'(\tfrac{1}{2}\Delta t'_{O'y'O'})\} = \{l_{O'y'}\},$$

$$\{l'(\tfrac{1}{2}\Delta t'_{O'x'O'})\} = \{l_{O'x'}\big/\sqrt{1-(v/V)^2}\},$$

where the numeric values of the physical quantities are indicated in the braces. In fact, until the equality of the physical quantity units in different reference systems is proven, the question of the content of the measured physical quantities remains open.

We note that the hardware components of groups $R$ and $R'$ cannot make a direct comparison of their kilometer and a moving kilometer by means of combining them at a certain moment in time before clock synchronization has been performed; therefore, the possibility of fixing the positions of moving elements at different points on the water surface at a given moment in time is not ensured.

## 6. Synchronization of the *R* and *R'* Group Clocks

Imagine that the readings of the group *R* clocks have been synchronized in such a manner that at a given moment in our time, *t*, they are identical according to our observation results. The group *R* hardware components can synchronize the clocks in two ways. First, they can transfer the readings of our clock to all the clocks on their barges at any moment in our time. And second, they can do this by obtaining information from us that their group is at rest on the water and that the velocity of the boats is identical in all directions. For example, in order to synchronize the clocks on the barges, one of which is located at the origin of coordinates, *O*, of the Σ system, while another is located on this system's *X* axis at a point with a coordinate of *x*, it is necessary to first send a boat from point *O* to point *x*, and once it reaches point *x*, to get it back. By taking time readings at point *O* at the moment that the boat is dispatched and at the moment that it comes back, the boat's time of movement between the origin of coordinates, *O*, and point *x* back and forth, $\Delta t_{OxO}$, can be determined. A boat carrying documents concerning the clock readings at the moment of boat dispatch must then again be dispatched from the origin of coordinates, *O*, to point *x*. Having received the document and the information concerning the fact that the boat trip time to a neighboring barge and back equals $\Delta t_{OxO}$, the hardware components on the barge at which the boat arrived, proceeding on the basis of the equality of the boat's velocity back and forth, establish a clock reading that exceeds the reading specified in the document they received by ½$\Delta t_{OxO}$. Thereafter, they in turn dispatch a boat to the barge on which the clock reading has not yet been synchronized. By continuing these actions, the hardware components on the barges of the group at rest, *R*, can synchronize all the *R* group clocks at the *X* axis. The clocks located at all the point in the Σ system can be synchronized in this manner.

We denote the group *R* time assigned by readings that are identical for all the group *R* barges using the symbol $t_R$.

Following clock synchronization, as outside observers, we see identical $t_R$ clock readings on all the barges of the group at rest on the water, *R*, at any moment in time. For purposes of notational convenience, we will assume that the *t* readings on our clock and the $t_R$ time readings on the group *R* clocks are always identical, i.e., $t = t_R$.

When the clocks on the barges of the group at rest, *R*, are synchronized, the slowing of the clock rate on each drifting barge can be recorded. To this end, it is sufficient to compare the changing reading of a $t_{r'}$ clock on a barge in motion, *r'*, to the readings of the $t_R$ clocks on the barges of the group at rest, *R*, that are located at different points on the water body surface. By pinpointing the position of the start and end of different rows of the groups in motion at a certain moment in time, the lengths of these rows can also be compared to one another and to the lengths of the group *R* rows. Following clock synchronization, the group *R* hardware components can measure the speed of movement of the group *R'* barges and the speed of movement of the auxiliary boats in any direction. It is not possible for the group *R'* hardware components to perform similar operations until this group's clocks have been synchronized.

Let's say the group *R'* clocks have also been synchronized in such a manner that, based on our observation results, the sameness of the clock readings on the different group *R'* barges is ensured at any moment in our time. We will denote the group *R'* time synchronized in this manner using the symbol $t'_{R'}$. Furthermore, we will presume that as a result of resetting to zero and due to random coincidences at the moment in time when the origins of coordinates, *O* and *O'*, of the Σ and Σ' systems are located at the same point, the clocks of all the barges in both groups will have a zero reading. It then follows from formula (8) that the group *R* and *R'* readings at any subsequent moment in time at any point on the water body will be linked to one another by the correlations

$$t'_{R'} = t_R \sqrt{1 - (v/V)^2} \text{ and } t_R = t'_{R'} \Big/ \sqrt{1 - (v/V)^2} \ . \qquad (18)$$

The $t_R$ and $t'_{R'}$ readings are not dependent upon the coordinates and ensure the absoluteness of simultaneity in both barge groups and in coordinate systems − Σ and Σ'. By absoluteness of simultaneity, we refer to the fact that if two events simultaneously occur at different locations within the Σ coordinate system at a moment in time of $t_R$, they also simultaneously occur in the Σ' system at a moment in time of $t'_{R'}$. Following clock synchronization in the *R* and *R'* groups, it becomes possible to fix the position of the ends of the barge rows in each of these groups at a certain moment in time, as well as to measure distances and lengths using not only the pseudoradar, but also the conventional method − by means of comparing them to one another or to proper length units.

Let's say that the lengths (the distances between barges) $l(½Δt_{αβα})$ and $l'(½Δt'_{α'β'α'})$ of the standards αβ and α'β', made up of a pair of barges, α and β, in the Σ coordinate system and a pair of barges, α' and β' in the Σ' system, are respectively used as distance measurement units in the groups at rest and in motion, *R* and *R'*.

We will presume that a speedboat requires time intervals of $Δt_{αβα}$ and $Δt'_{α'β'α}$ in order to traverse each of these distances back and forth, so that the numeric values of these intervals are standardized and equal one another. If the $l(½Δt_{αβα})$ distance measurement unit of the standard at rest has the same denomination as the $l'(½Δt'_{α'β'α'})$ simulated distance measurement unit of the standard in motion, for example, if both units are called a kilometer, then as is clear from formulas (12) and (17), the simulated and conventional lateral kilometers equal one another, while according to our data, the simulated moving longitudinal kilometer is $1/\sqrt{1-(v/V)^2}$ times shorter than the resting conventional kilometer.

Taking into account the fact that the distance we measured between points $O'$ and $x'$, $l_{O'x'}$, equals $x-vt_R$, where $x$ – the coordinate of point $x'$ in the $\Sigma$ system, and assuming that $l'(½\Delta t'_{O'x'O'}) = x'$, then from the right-hand equality of correlation (17), we obtain the direct transformation of the coordinates

$$x' = (x - vt_R)/\sqrt{1-(v/V)^2}. \qquad (19)$$

Returning to the synchronization of the $R'$ group clocks, we note that the group $R'$ hardware components can only technically perform clock synchronization of this type by means of directly transferring the group $R$ clock readings at a certain moment in time to the barges of their group, or through the use of speedboats. In the latter instance, the hardware components of the group $R'$ barges must be furnished with information on precisely which $R'$ group is in motion, during which it is moving so that the simulated speed of movement of the water relative to the $\Sigma'$ coordinate system, $v'$, is aimed in the direction opposite that of the $X'$ axis. If we know the rule for converting the velocity that we observe from the outside, $v$, into simulated velocity, $v'$, and the rule for converting the $V$ velocity into $V'$ simulated velocity, we can then transfer $v'$ and $V'$ simulated velocity information to the hardware components operating on the simulated time scale. The latter can be expressed in simulated distance and simulated time units. Knowing the rule expressed by formula (19) for converting the $x$ coordinate into the $x'$ coordinate, regarding the ratio $-x'/t'_{R'}$ as the point $O$ velocity (the origin of coordinates of the $\Sigma$ system) in the $\Sigma'$ system, and equating $x$ to zero, then from formulas (18) and (19), we obtain

$$v' = v/(1 - v^2/V^2). \qquad (20)$$

Substituting first $x = -Vt_R$, then $x = Vt_R$ in formula (19), we find the absolute value of the $-x'/t'_{R'}$ boat velocity in the $\Sigma'$ system with the current in the first case and the absolute value of the $x'/t'_R$ boat velocity against the current in the second case. Denoting $-x'/t'_{R'}$ for movement with the current through $V_1$ and $x'/t'_{R'}$ for movement against the current through $V_2$, and taking $t'_{R'}$ from the left-hand side of formula (18), we obtain:

$$V'_1 = (V + v)/(1 - v^2/V^2); \qquad (21)$$
$$V'_2 = (V - v)/(1 - v^2/V^2). \qquad (22)$$

Introducing the notation

$$V' = V/(1 - v^2/V^2)$$

and dividing the $v'$ velocity presented in formula (20) by $V'$, we obtain

$$v'/V' = v/V. \qquad (23)$$

Taking formulas (20) and (23) into account, equalities (21) and (22) can be presented in the form

$$V'_1 = V' + v' \qquad (24)$$
$$V'_2 = V' - v'. \qquad (25)$$

The $V'_1$ and $V'_2$ velocities represent the simulated boat velocities in the $\Sigma'$ system with the water current and against the current. The physical significance of the $V'$ quantity is clear from formulas (24) and (25). This quantity represents the boat velocity relative to the water surface, expressed in simulated group $R'$ distance (length) and time units.

Having received information on the $v'$ and $V'$ velocities, the group $R'$ hardware components, using formulas (24) and (25), can calculate the $V'_1$ and $V'_2$ velocities for the boats with the current and against the current. Then, by dispatching a boat from barge $r'$ to point $O'$ and to point $x'$, they can also transfer the clock reading from barge $r'$ to the barge at point $x'$, having added the $V'_2 \Delta t'_{O'x'O'}$ quantity to this reading.

Using formula (23) and bearing in mind the independence of time upon the coordinate, which makes it possible to group the symbols using simple algebraic rules, we can write formula (20) in the form

$$v = v'(1 - v'^2/V'^2). \tag{26}$$

Substituting $v$ and $t_R = t'_{R'}/\sqrt{1-(v/V)^2}$ in formula (19), performing direct algebraic transformations (they are permissible due to the independence of time upon the coordinate), and taking formula (23) into account, we obtain the transformation

$$x = (x' + v't'_{R'})\sqrt{1-(v'/V')^2}, \tag{27}$$

where $V'$ and $v'$ appear in the expression under the radical.

Using formula (19) and the left-hand side of formula (18), the direct coordinate and time transformations can be written in the form

$$x' = (x - vt_R)/\sqrt{1-(v/V)^2}, \; y' = y, \; t'_{R'} = t_R\sqrt{1-(v/V)^2}, \tag{28}$$

Using the right-hand side of formula (18) and formula (27), the inverse transformations can be written

$$x = (x' + v't'_{R'})\sqrt{1-(v'/V')^2}, \; y' = y \; t_R = t'_{R'}/\sqrt{1-(v'/V')^2}. \tag{29}$$

Transformations (28) and (29) for the $x$ and $x'$ coordinates and the $t_R$ and $t'_{R'}$ time values are asymmetrical. If the group $R$ and $R'$ hardware components perceive their proper longitudinal kilometer and their proper second as key, they then perceive the kilometer and second of the other group as erroneous. In this instance, the group $R$ hardware components regard the group $R'$ kilometer as too short and the group $R'$ second as extremely protracted. At the same time, the group $R'$ hardware components regard the group $R$ kilometer as too long and the second as truncated. In general, the hardware components of the group at rest regard the geometric dimensions of the moving objects made up of the barges in motion, as well as the passage of time on the individual barges in motion, like they regard the observers or the instruments in a given coordinate system of the special theory of relativity. However, the group $R'$ hardware components regard the dimensions of the objects made up of the barges at rest and the passage of time thereon in a manner opposite the one in which the instruments of a coordinate system of the special theory of relativity do this.

## 7. Simulation of the Symmetry of Relativistic Effects

In order to derive the symmetry of the coordinate and time transformations used in the special theory of relativity, it is necessary to simulate $R'$ group ($\Sigma'$ system) clock synchronization in such a manner that the speeds of movement of a boat from the origin of coordinates, $O'$, to point $x'$ and back to the origin of coordinates, $O'$, are identical. During the time over the course of which a boat moving at a velocity of $V - v$ travels from the origin of coordinates, $O'$, to point $x'$, $\Delta t_{O'x'}$, the $t$ reading on our clock is increases by $\Delta t = l_{O'x'}/(V - v)$, while the $t'_{R'}$ reading on the group $R'$ clocks, which run $1/\sqrt{1-(v/V)^2}$ times more slowly than our clock, increases by $\Delta t' = l_{O'x'}\sqrt{1+v/V}/(V\sqrt{1-v/V})$.

If a boat is dispatched from the origin of coordinates, $O'$, of the $\Sigma'$ system to point $x'$, then returns from point $x'$ to the origin of coordinates, $O'$, the reading of our clock will increase by $\dfrac{2l_{O'x'}}{V(1-v^2/V^2)}$ during this boat's voyage according to formula (13). During this period of time, the readings on the group $R'$ barges will increase by a value that is $1/\sqrt{1-(v/V)^2}$ times less and that equals $2l_{O'x'}/(V\sqrt{1-(v/V)^2})$. The equality of the boat's velocities in opposite directions within the $\Sigma'$ system can be achieved if the difference in the group $R'$ clock reading at point $x'$ at the moment that the boat arrives there and the clock reading at the origin of the coordinate system at the moment that the boat is dispatched from it is not $l_{O'x'}\sqrt{1+v/V}/(V\sqrt{1-v/V})$, but rather half the value of $2l_{O'x'}/(V\sqrt{1-(v/V)^2})$, i.e., $l_{O'x'}/(V\sqrt{1-(v/V)^2})$. I.e., in order to achieve the equality of boat's velocities in opposite directions, the $t''_{R'}$ clock reading at a point with a coordinate of $x'$ must be less than the $t'_{R'}$ reading for the difference in the $l_{O'x'}\sqrt{1+v/V}/(V\sqrt{1-v/V})$ and $l_{O'x'}/(V\sqrt{1-(v/V)^2})$ values. Taking into account the fact that $l_{O'x'} = x'\sqrt{1-(v/V)^2}$, this difference equals $l_{O'x'}v/(V^2\sqrt{1-(v/V)^2})$, i.e.,

$$t'_{R'} - t''_{R'} = \frac{l_{O'x'}v}{V^2\sqrt{1-(v/V)^2}},$$

whence

$$t''_{R'} = t'_{R'} - \frac{l_{O'x'}v}{V^2\sqrt{1-(v/V)^2}} \qquad (30)$$

and

$$t'_{R'} = t''_{R'} + \frac{l_{O'x'}v}{V^2\sqrt{1-(v/V)^2}} \qquad (31)$$

Using formulas (30) and (19), as well as the left-hand side of formula (18), and bearing in mind that $l_{O'x'} = x - vt_R$, we obtain $t''_{R'} = (t_R - xv/V^2)/\sqrt{1-(v/V)^2}$, which, together with formula (19), yields the transformations

$$x' = (x - vt_R)/\sqrt{1-(v/V)^2} \; ; \; y' = y \; ; \; t''_{R'} = (t_R - xv/V^2)/\sqrt{1-(v/V)^2} \qquad (32)$$

The transformations in formula (32) differ radically from those in formula (28).
We will call the $t''_{R'}$ time the double simulated time and the quantities expressed through this time the double simulated quantities. From the transformations in formula (32), it is possible to find the $x'/t''_{R'}$ ratio under the condition of $x = 0$. This ratio represents the double simulated velocity, $v''$, of the origin of coordinates of the $\Sigma$ system within the $\Sigma'$ system and equals the $v$ velocity, i.e.,

$$v'' = v.$$

The equality of the $V$ boat velocity to the $V''$ double simulated boat velocity in any direction follows from the condition of the equality of the boat velocities en route back and forth, as well as from the condition of the equality of the boat velocity there and the boat velocity back. Using the $v'' = v$ and $V'' = V$ equalities obtained, it is easy to find the inverse transformations from the transformations in formulas (28) and (31), which take the form

$$x = (x' + v''t''_{R'})/\sqrt{1-(v''/V'')^2}\ ;\ y = y';\ t_R = (t''_{R'} + x'v''/V''^2)/\sqrt{1-(v''/V'')^2}\ . \qquad (33)$$

The transformations obtained in formulas (32) and (33) for our two-dimensional case are indistinguishable from the Lorentz transformations to notational accuracy. The transformations are symmetrical. The group $R$ and $R'$ hardware components, which perceive their proper longitudinal kilometer and their proper second as key, perceive the geometric dimensions of the other group's objects, including the dimensions of the kilometer standard, as abbreviated and the time as slow.

## 8. Addition of Velocities

The result of the addition of the simulated velocities is obvious, since it follows from the transformations in formulas (32) and (33), which are indistinguishable from the Lorentz transformations. If, for example, a certain watercraft moves at a velocity of $u$ within the $R$ group $\Sigma$ coordinate system in the direction of the $X$ axis and the equation $x = ut_R$ describes the dependence of its $x$ coordinate upon time, then by substituting $ut_R$ in place of $x$ in the first transformation in formula (32), we obtain $x' = (u-v)t_R/\sqrt{1-(v/V)^2}$. By substituting $x = ut_R$ in the last transformation in formula (32), we obtain $t''_{R'} = (1-uv/V^2)t_R/\sqrt{1-(v/V)^2}$. Dividing the $x'$ value of the watercraft we are examining by $t''_{R'}$, we obtain the double simulated velocity, $u''$, which can be expressed by the formula

$$u'' = \frac{u-v}{1-uv/V^2}.$$

If the watercraft move in accordance with the law $x = -ut_R$, i.e., in the direction opposite that of the $X$ axis, the velocity addition formula then takes the form

$$u'' = \frac{u+v}{1+uv/V^2}.$$

It is apparent from the latter formula that the watercraft's double simulated velocity cannot exceed the $V$ velocity, and when the $u$ and $v$ velocities approach $V$, it is not $2V$ that is approached, but rather $V$.

## 9. Simulation of the Simplest Effects of Noninertial Bodies

The simulation model we are proposing also makes it possible to examine the simplest effects on noninertial barges and within noninertial barge groups.

Let's say that each individual barge group simulates a solid physical body. The immutability of the distances between the component parts of the solid body, fixed in its own coordinate system, simulate the maintenance of the radar distance between the barges in each group.

We will examine two barge groups at rest, which, located a great distance from one another, simultaneously get under way and accelerate in accordance with identical programs along the line on which they are located according to our clock and according to the synchronized clocks of the barge groups at rest. In this case, according to our observations, the longitudinal distances between the barges in each of the two groups moving one after the other begin to shorten after a certain time. Here, the instruments on the barges will maintain the immutability of the radar distances within each group. However, according to our observation results, the distance between the barge groups remains immutable, first, due to the sameness of the acceleration programs, and second, since the tracking and maintenance of the distances between the barges are only performed within each group. I.e., from our point of view as outside observers, each group is contracted in the direction of movement, while the distance between the groups remains the same as before. After acceleration comes to an end, if the group instruments measure the distance between the groups using boats that make trips back and forth, or using lengths standards, they will then find that the groups have drifted apart and that the distance between them has increased. This effect is presently called the Bell paradox, although it was described before Bell, in particular, by D. V. Skobeltsyn [5].

The increase in the distance between the groups recorded by the instruments is due to the contraction of the groups, as well as the length standard that pertains to them, in the direction of movement and has a purely metrological relative nature.

After a certain time, if the barge groups begin to perform a reverse operation and simultaneously begin to decelerate, the result of this deceleration will then be dependent upon group clock synchronization.

After acceleration comes to an end and before deceleration begins, if the instruments within the groups do not resynchronize the clocks and simultaneously begin to decelerate in our time, $t$, and in that same time, $t_R$ (as well as simultaneously in single stimulated time, $t'_{R'}$), the reverse process will occur in such a manner that the groups and the length standards that pertain to them, following the completion of deceleration and standstill, will again be expanded, and the groups will return to the original state. Due to the expansion of the length standards, the instruments will record a decrease in distance to the initial (starting) value. This will occur at an immutable actual distance between the groups.

If, however, the instruments of a pair of groups do resynchronize the clocks "according to Einstein" and the groups simultaneously begin to decelerate in double simulated time, $t''_{R'}$, then according to our clock, they begin to decelerate at different times (the rear group of the pair begins to decelerate earlier that the front group). In this instance, despite the expansion of the groups and the standards, after the groups of barges stop in the water, the instruments record a sequent increase in the distance between the groups, since, as it is easy to demonstrate, the increase in the distance between the groups due to the actual difference in the times that they begin to decelerate exceeds the expansion of the groups and the length standards by $1/\sqrt{1-(v/V)^2}$ times.

The Bell paradox will be examined in greater detail in our book.

## 10. Simulated "Space-Time"

In deriving the transformations in formulas (32) and (33), we did, as a matter of fact, also simulate pseudo-Euclidean space-time, since these transformations ensure the invariance of the space-time interval in the Σ and Σ' systems, as well as in any other systems associated with groups of barges in motion at different velocities. It is clear that this "space-time" has nothing at all to do with enigmatic multidimensional worlds and is an elementary mathematical construct that pertains more to the formalization of the measurement errors caused by the failure to take the presence of an aquatic environment into account than to the behavioral features of the barges on the water surface. But this is a separate topic of discussion. We will dwell on this subject in greater detail in the book on which we are presently working.

**Conclusion**

The special theory of relativity is closely linked to philosophical suppositions, and many of the problems that arise over the course of interpreting its physical content are philosophical in nature. Interpretational problems are not to be solved by constraining philosophers from solving philosophical problems in modern physics. The total disagreement in evaluating the content of the concepts used in the special theory of relativity can serve as evidence of this, during which it is not so much philosophers that create this disagreement as physicists themselves. Hermann Weyl probably foresaw something similar when he wrote: "Despite the disheartening leapfrog of philosophical systems, we cannot refuse to address it unless we want knowledge to be transformed into senseless chaos" [6].

What is the essence of kinematic phenomena? Is Lorentz contraction a real or apparent phenomenon? Is four-dimensional space-time an objective reality or a mathematical formalism? Physicists give not so much categorical answers to these and other questions as answers that mutually exclude one another.

The special theory of relativity is very simple and does not involve any problems other than the problems of its interpretation. It can be described in the simplest language and using the simplest examples from our everyday life.

The simulation presented in this article quite dramatically demonstrates the simplicity of the special theory of relativity and its "earthiness". It is not difficult to see that the possibility of using a four-dimensional formalism that does not differ from Minkowski formalism in this model issues from a simulation that yield Lorentz transformations.